\documentclass[prl,aps,reprint,twocolumn,showpacs,superscriptaddress]{revtex4-1}
\usepackage{graphicx,color,bm}
\usepackage[english]{babel}
\usepackage{amsmath,amssymb}
\usepackage{dsfont}
\usepackage{subfigure}

\newcommand{\green}[1]{{\textcolor{black}{#1}}}
\newcommand{\red}[1]{{\textcolor{black}{#1}}}

\begin{document}

\title{Blurring the boundaries between topological and non-topological phenomena in dots}

\author{Denis R. Candido}
\affiliation{Instituto de F\'{\i}sica de S\~ao Carlos, Universidade de S\~ao Paulo, 13560-970, S\~ao Carlos, S\~ao Paulo, Brazil}
\author{M. E. Flatt\'e}
\affiliation{Department of Physics and Astronomy and Optical Science and Technology Center, University of Iowa, Iowa City, Iowa 52242, USA}
\author{J. Carlos Egues}
\affiliation{Instituto de F\'{\i}sica de S\~ao Carlos, Universidade de S\~ao Paulo, 13560-970, S\~ao Carlos, S\~ao Paulo, Brazil}
\affiliation{International Institute of Physics, Federal University of Rio Grande do Norte, 59078-970, P. O. Box 1613, Natal, Brazil}
\date{\today}

\begin{abstract}
We investigate the electronic and transport properties of topological and trivial InAs$_{1-x}$Bi$_x$ quantum dots (QDs). By considering the rapid band gap change within  valence band anticrossing theory for  InAs$_{1-x}$Bi$_x$,  we show that Bi-alloyed quantum
wells become $\sim 30$~meV gapped 2D topological insulators for well widths $d>6.9$nm $(x = 0.15)$
and obtain the $\boldsymbol{k.p}$ parameters of the corresponding Bernevig-Hughes-Zhang (BHZ) model. We
analytically solve this model for cylindrical confinement via modified Bessel functions. For 
non-topological dots we find ``geometrically protected" discrete helical edge-like states, i.e., Kramers pairs with spin-angular-momentum locking, in stark contrast with ordinary InAs QDs. For a conduction window with four edge states, we  find that the two-terminal  conductance ${\cal G}$ vs. the QD radius $R$ and the gate $V_g$ controlling its levels shows a double peak at $2e^2/h$ for both topological and trivial QDs. In contrast, when bulk and edge-state Kramers pairs coexist and are degenerate, a single-peak resonance emerges.  Our results blur the boundaries between topological and non-topological phenomena for conductance measurements in small systems such as QDs. 
Bi-based BHZ QDs should also prove important as hosts to edge spin qubits.
\end{abstract}

\maketitle

{\it Introduction.---} Topological Insulators (TIs) are a new class of materials having the unusual property of being an insulator in bulk with robust gapless helical states localized near their edges (2D TIs) and surfaces (3D TIs)~\cite{graphene,science2006,science2007}. Following these pioneering works, a few other TI proposals~\cite{prbsig,prl-sush,prl-kai, inasgasb,3dti1,3dti2,insbstrain} have been put forward with some experimental support~\cite{inasgasbexp,fabrizio}. More recently, topological QDs with cylindrical confinement have been investigated~\cite{tiqd1,tiqd2,tiqd3,tiqd4,tiqd5,tiqd6,tiqd7,tiqd8,tiqd9,tiqd10,tiqd11,tiqd12}. Their spectra feature discrete helical edge states protected against non-magnetic scattering and showing spin-angular-momentum locking. These states are potentially important for spintronics~\cite{tiqd2,tiqd3}, quantum computation and other quantum technologies~\cite{tiqd1,tiqd4,tiqd4,tiqd5}.
\begin{figure}[t!]
\centerline{\resizebox{3.45in}{!}{\includegraphics{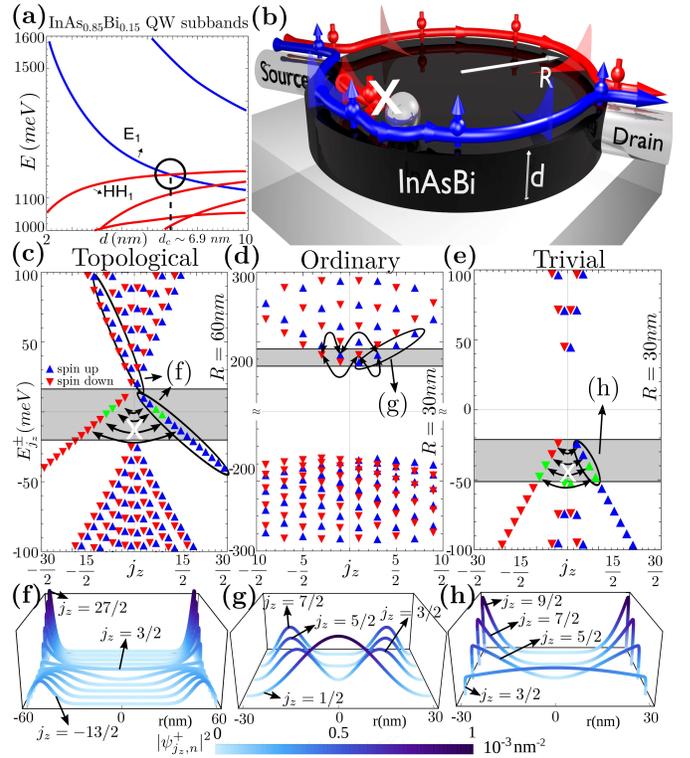}}}
\caption{\label{fig1}(a) InAs$_{0.85}$Bi$_{0.15}$ QW subbands vs. the well thickness $d$ and (b) schematic of a cylindrical QD with helical edge states. Energy levels vs. the total angular momentum $j_z$ for (c) a topological InAs$_{0.85}$Bi$_{0.15}$  QD with $R=60$~nm, (d) an ordinary InAs QD with $R=30$~nm and (e) a trivial InAs$_{0.85}$Bi$_{0.15}$ QD with $R=30$~nm. The curved arrows denote forbidden and allowed transitions. (f), (g) and (h): Modulus square of the spin up wave functions $|\psi_{j_z,n}^+|^2$ for the edge states grouped by the ellipses in (c), (d) and (e). 
}
\end{figure}

\begin{figure}[t]
\centerline{\resizebox{3.5in}{!}{
\includegraphics{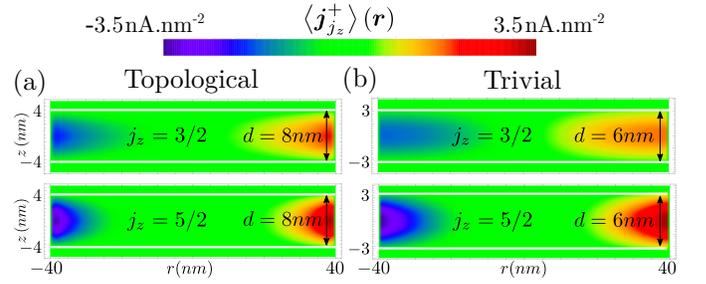}}}
\caption{Spin-up circulating currents for topological (a) and trivial (b) edge states $j_z=\frac{3}{2}$ and $j_z=\frac{5}{2}$, see green triangles within the gray area in Figs.~\ref{fig1}(c) and \ref{fig1}(e).  The topological and trivial circulating currents are essentially the same. The horizontal white lines  delimit the soft-wall QW barriers.}
\label{fig2}
\end{figure}

\begin{figure}[t]
\centerline{\resizebox{3.6in}{!}{
\includegraphics{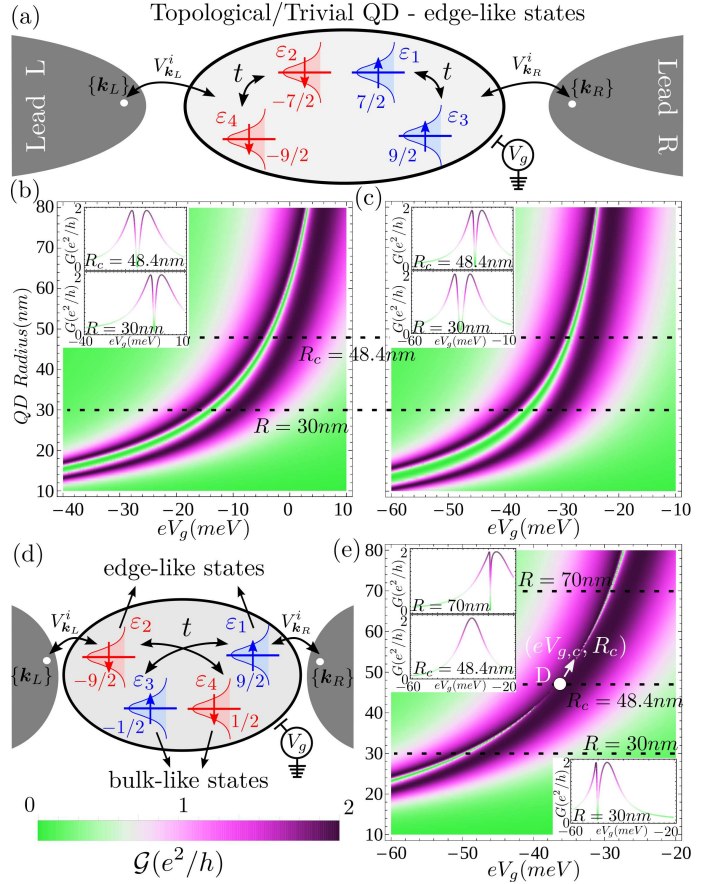}}}
\caption{(a) Schematic QD Hamiltonian for the four topological and trivial edge states with $j_z=\pm 7/2$ and $j_z=\pm 9/2$. QD conductance ${\cal G}$ at $T=0$K for the topological (b) and trivial (c) edge states in Fig.~\ref{fig3}(a). (d) Same as (a) for the coexisting $j_z=\pm 1/2$ bulk and $j_z=\pm 9/2$ edge states and the corresponding ${\cal G}$ for the trivial case (e).} 
\label{fig3}
\end{figure}

In this work, we first predicts that InAs$_{1-x}$Bi$_{x}$/AlSb quantum wells (QWs) become 2D topological insulators for well widths $d> 6.9$ nm and $x=0.15$, with large inverted subband gaps $\sim 30$ meV ($> k_BT$) that should enable room temperature applications~[Fig.\ref{fig1}(a)]. \red{Using the valence band anticrossing theory and the $\boldsymbol{k.p}$ approach, we also determine the effective parameters of the Bernevig-Hughes-Zhang (BHZ) model for our Bi-based wells. Then we  define cylindrical quantum dots (QDs) by confining the BHZ model with soft and hard walls, and perform {\it analytical} electronic structure calculations for the QD energy levels and wave functions (Fig.~\ref{fig1}) both in the topological and trivial regimes.} Surprisingly, we find that the {\it trivial} QDs have {\it geometrically protected helical edge states} [Figs.~\ref{fig1}(e) and \ref{fig1}(h)] with spin-angular-momentum locking similar to topological QDs, Figs.~\ref{fig1}(c) and \ref{fig1}(f), \red{and in contrast to ordinary QDs, Figs.~\ref{fig1}(d) and \ref{fig1}(g).} These trivial helical edge-like states occur in a wide range of QD radii and lie outside the \red{BHZ bandgap. We have also calculated the circulating currents~\cite{michael1,michael2} (Fig.~\ref{fig2}) and the two-terminal QD conductance ${\cal G}$ within linear response~\cite{meir-wingreen}, Fig.~\ref{fig3}. Interestingly, topological and trivial QDs exhibit similar transport properties, e.g., the conductance of QDs with two Kramers pairs of edge states show double-peak resonances at ${\cal G}=2 e^2/\hbar$, separated by a dip due to destructive interference in both the topological and trivial regimes. When bulk and edge-state Kramers pairs coexist {\it and} are degenerate, both regimes show a single-peak resonance also at ${\cal G}=2 e^2/\hbar$. Our findings blur the boundary between topological and non-topological QDs as for \green{the appearance of protected helical edge states} and conductance measurements.}

\paragraph*{New 2D Topological Insulator: {\rm InAs$_{0.85}$Bi$_{0.15}$/AlSb}.---} The response of the electronic structure of InAs to the addition of the isoelectronic dopant Bi~\cite{MBEInAsBi} is well described within valence band anticrossing theory~\cite{Alberi2007}. Bi provides a resonant state within the valence band (complementary to the resonant state in the conduction band generated in the dilute nitrides such as GaAs$_{1-x}$N$_{x}$) which strongly pushes up the valence band edge of InAs as Bi is added. The  small band gap of InAs allows it to close for approximately 7.3\% of Bi~\cite{MBEInAsBi}, and for inversion of the conduction and valence bands similar to HgTe for larger Bi percentage. We determine the electronic states of a InAs$_{1-x}$Bi$_{x}$/AlSb QW grown on a GaSb substrate~(SM, Sec.~I) within a superlattice electronic structure calculation implemented within a fourteen bulk band basis~\cite{LOF} and obtain the zone-center ($\Gamma$ point, Fig.~\ref{fig1}(a)) quantum well states. From those we derive momentum matrix elements and the other parameters of the BHZ Hamiltonian. We obtain a crossing between the lowest conduction subbands $\left|E_{1}\pm\right\rangle$ and the highest valence subbands $\left|HH_{1}\pm\right\rangle$ at the critical well thickness $d_c=6.9$nm. This crossing characterizes a topological phase transition between an ordinary insulator ($d<d_c$) and a 2D TI ($d>d_c$) \red{with an inverted gap $\sim 30$~meV, Fig.~\ref{fig1}(a).}

\paragraph*{ Model Hamiltonian for a cylindrical dot.---} \red{We consider the BHZ Hamiltonian describing} the low-energy physics of the $\left|E_{1}\pm\right\rangle$ and $\left|HH_{1}\pm\right\rangle$ subbands,	 
\begin{equation}
{\cal H}\left(\mathbf{k}\right)=\left(\begin{array}{cc}
H\left(\mathbf{k}\right) & 0\\
0 & H^{*}\left(-\mathbf{k}\right)
\end{array}\right),
\label{bhz}
\end{equation}
where $H\left(\mathbf{k}\right)=(C-D\mathbf{k}^2)\boldsymbol{1_{2\times2}}+\boldsymbol{d}\cdot\boldsymbol{\sigma}$ and $\boldsymbol{d}(\mathbf{k})=\left(\thinspace Ak_{x},-Ak_{y},\thinspace M-B\mathbf{k}^2\right)$. Here, $\mathbf{k}$ is the in plane wave vector and $\boldsymbol{\sigma}$ are the Pauli matrices describing the pseudo-spin space. The parameters $A,\thinspace B,\thinspace C,\thinspace D,\thinspace M$, calculated within \red{a superlattice $\boldsymbol{k.p}$ electronic structure calculation~\cite{LOF}}, depend on the QW thickness $d$ and are given in Tab.~(S1) of the SM for $d=6$~nm ($x=0.15$) and $d=8$~nm ($x=0.15$). \red{We define our QDs by adding to Eq.~(\ref{bhz})} the in-plane cylindrical confinement~\green{\cite{tiqd1,tiqd2,tiqd3,tiqd4,tiqd5,tiqd6,tiqd7,tiqd8,tiqd9,tiqd10,tiqd11,tiqd12,tiqd13}}
\begin{equation}
V_{c}=\left(\begin{array}{cc}	 
V\left(r\right)\sigma_{z} & 0\\
0 & V\left(r\right)\sigma_{z}
\end{array}\right),\: V\left(r\right)=\begin{cases}
0 & r<R\\
M_{O}-M & r>R.\\ 
\end{cases}
\label{conf}
\end{equation}
The soft wall confinement in Eq.~(\ref{conf})~\green{\cite{poli}} has equal strength barriers $\left(M_O>0\right)$ for electrons and holes. Here we focus on the hard wall case ($M_O\rightarrow\infty$) as it is simpler analytically. In the SM we discuss the soft wall case, which qualitatively shows the same behavior. 

\paragraph*{\red{Analytical QD eigensolutions.}} To solve $[{\cal H}\left(\mathbf{k}\right) + V_c] \psi= \varepsilon\psi$, we make $k_{x}\rightarrow-i\partial_{x}$ and $k_{y}\rightarrow-i\partial_{y}$ or 
\begin{eqnarray}
k_x\pm ik_y & \rightarrow & -i\thinspace e^{\pm i\theta}\left(\frac{\partial}{\partial r}\pm\frac{1}{r}\frac{\partial}{\partial\theta}\right)\label{kplus},\\
\mathbf{k}^2 & \rightarrow & -\left(\partial_{r}^{2}+\frac{1}{r}\partial_{r}+\frac{1}{r^{2}}\partial_{\theta}^{2}\right)\label{ksquare},
\end{eqnarray}
\red{in polar coordinates. By imposing that $\psi(r,\theta)=0$ at $r=R$}, we obtain the transcendental equation for all the quantized eigenenergies and the corresponding analytical expressions for the wave functions 
\begin{widetext}
\begin{align}
\frac{\lambda_{-}^{2}\left(E_{j_{z},n}^{\pm}\right)-\frac{E_{j_{z},n}^{\pm}-C-M}{D+B}}{\lambda_{-}\left(E_{j_{z},n}^{\pm}\right)}\frac{I_{j_{z}\mp\frac{1}{2}}\left[\lambda_{+}\left(E_{j_{z},n}^{\pm}\right)R\right]}{I_{j_{z}\mp\frac{1}{2}}\left[\lambda_{-}\left(E_{j_{z},n}^{\pm}\right)R\right]}=\frac{\lambda_{+}^{2}\left(E_{j_{z},n}^{\pm}\right)-\frac{E_{j_{z},n}^{\pm}-C-M}{D+B}}{\lambda_{+}\left(E_{j_{z},n}^{\pm}\right)}\frac{I_{j_{z}\mp\frac{3}{2}}\left[\lambda_{+}\left(E_{j_{z},n}^{\pm}\right)R\right]}{I_{j_{z}\mp\frac{3}{2}}\left[\lambda_{-}\left(E_{j_{z},n}^{\pm}\right)R\right]},\label{eq5}\qquad\qquad\qquad\\
\psi_{j_{z},n}^{\pm}\left(r,\theta\right)=\frac{Ne^{ij_{z}\theta}}{\sqrt{2\pi}}\left[\begin{array}{c}
\left(I_{j_{z}\mp\frac{1}{2}}\left(\lambda_{+}\left(E_{j_{z},n}^{\pm}\right)r\right)-\frac{I_{j_{z}\mp\frac{1}{2}}\left(\lambda_{+}\left(E_{j_{z},n}^{\pm}\right)R\right)}{I_{j_{z}\mp\frac{1}{2}}\left(\lambda_{-}\left(E_{j_{z},n}^{\pm}\right)R\right)}\thinspace I_{j_{z}\mp\frac{1}{2}}\left(\lambda_{-}\left(E_{j_{z},n}^{\pm}\right)r\right)\right)e^{\mp i\frac{\theta}{2}}\\
\frac{\left(D+B\right)\lambda_{+}^{2}\left(E_{j_{z},n}^{\pm}\right)-E_{j_{z},n}^{\pm}+C+M}{\pm iA\lambda_{+}\left(E_{j_{z},n}^{\pm}\right)}\left(I_{j_{z}\mp\frac{3}{2}}\left(\lambda_{+}\left(E_{j_{z},n}^{\pm}\right)r\right)-\frac{I_{j_{z}\mp\frac{3}{2}}\left(\lambda_{+}\left(E_{j_{z},n}^{\pm}\right)R\right)}{I_{j_{z}\mp\frac{3}{2}}\left(\lambda_{-}\left(E_{j_{z},n}^{\pm}\right)R\right)}\thinspace I_{j_{z}\mp\frac{3}{2}}\left(\lambda_{-}\left(E_{j_{z},n}^{\pm}\right)r\right)\right)e^{\mp i\frac{3\theta}{2}} \label{eq6}
\end{array}\right].
\end{align}
\end{widetext}
Here $I_{j_z}(\lambda_{\pm}\left(E_{j_z,n}^{\sigma}\right) r)$ is the modified Bessel's function of the first kind, $N$ a normalization factor and $\lambda_{\pm}^{2}\left(E_{j_z,n}^{\sigma}\right)=-F\pm\sqrt{F^{2}-Q^{2}}$ with $F=\frac{1}{2}\left(\frac{A^{2}}{\left(D+B\right)\left(D-B\right)}-\frac{E_{j_z,n}^{\sigma}-C-M}{D+B}-\frac{E_{j_z,n	}^{\sigma}-C+M}{D-B}\right)$ and $Q^2=\left(\frac{E_{j_z,n}^{\sigma}-C+M}{D-B}\right)\left(\frac{E_{j_z,n}^{\sigma}-C-M}{D+B}\right)$. \red{The $\pm$ signs in \red{Eqs.}~(\ref{eq5}) and (\ref{eq6}) label the ``spin" subspaces in the BHZ model (i.e., its two 2x2 blocks)~\cite{bhzspin}, and arise \red{as} the Time Reversal Symmetry (TRS) operator $\Theta=-i\sigma_{y}\otimes\boldsymbol{1_{2\times2}}K$ commutes with ${\cal H}\left(\mathbf{k}\right)$ in Eq.~(\ref{bhz}).} The $\psi^\pm$ states in (\ref{eq6}) form a Krammers pair,  \red{i.e.,} $\Theta\psi_{j_z,n}^{+}(r,\theta)=\psi_{-j_z,n}^{-}(r,\theta)$. \red{The} quantum number $j_z$ corresponds to the z-component of the total angular momentum ${\cal J}_{z}=-i\hbar \partial_{\theta}+\hbar\sigma_z\otimes(\tau_0-\frac{\tau_z}{2})$ \red{that obeys} ${\cal J}_z \psi_{j_{z},n}^{\pm}(r,\theta)=\hbar j_z \thinspace \psi_{j_z,n}^{\pm}(r,\theta)$\red{, $j_z=\pm\frac{1}{2},\pm\frac{3}{2},...$. Incidentally, $j_z$ also denotes the parity of the QD states defined via the} inversion symmetry \red{operator ${\cal I}$} $\left(r,\theta\right)\rightarrow \left(r,\theta+\pi\right)$, \red{satisfying} ${\cal I}\psi_{j_{z},n}^{\pm}\left(r,\theta\right)=\left(-1\right)^{j_z \mp \frac{3}{2}}\psi_{j_{z},n}^{\pm}\left(r,\theta\right)$. \red{Both ${\cal J}_{z}$ and ${\cal I}$ commute with the QD Hamiltonian. The quantum number $n$ arises from the radial confinement of the dot; we index our energy spectrum such that for each $j_z$ and $\sigma (=\pm)$ $n=1,2,3...$ ($n=-1,-2,-3,...$ ) for positive (negative) energies.} 

In Figures~\ref{fig1}(c), \ref{fig1}(d) and \ref{fig1}(e), we plot the InAs$_{1-x}$Bi$_x$ QD energy levels [Eq.~(\ref{eq5})] for {\it topological} ($x=0.15$, $d=8$~nm, $R=60$~nm), {\it ordinary} ($x=0$, $d=6$~nm, $R=30$~nm) and {\it trivial} ($x=0.15$, $d=6$~nm, $R=30$~nm) cases respectively. \green{The {\it ordinary} InAs QD with its non-inverted large gap is considered here for \green{comparison~(SM, Sec. IV)}}. In contrast to the BHZ model with one (or two) interface(s), where both the edge and bulk dispersions are continuous, here we have discrete energy levels.

\paragraph*{Geometrically protected trivial helical edge states.---} 
Surprisingly, we find for the trivial QD [Figs.~\ref{fig1}(e), \ref{fig1}(h)] spin resolved single Kramers pairs edge states with spin-angular-momentum locking similarly to the topological QD [Figs.~\ref{fig1}(c), \ref{fig1}(f)], except that here these edge-like states lie outside the gap. The number of these protected trivial helical edge-like states within the gray area is proportional to the modulus of the BHZ particle-hole asymmetry \red{term $B$} and they appear in the valence (conduction) subspace for $B<0$ ($B>0$). In contrast to the topological QD, these helical edge states are geometrically protected by the QD confinement, which prevents the coexistence of bulk-like and edge-like valence states within the gray area in Fig.~1(e). In the SM we show that our results hold for a wide range of QD radii and other BHZ parameters, in particular those of HgTe/CdTe QDs~\green{\cite{science2006,ewelina,zhangreview}.}


In contrast, ordinary cylindrical InAs QDs defined from InAs wells with parabolic subbands do not have protected edge-like states [Fig.~\ref{fig1}(d), \ref{fig1}(g)]. These QDs have the degeneracies $E_{j_z\mp\frac{1}{2},n}^{E_1 \pm}=E_{-j_z\pm\frac{1}{2},n}^{E_1 \pm}$ and $E_{j_z\mp\frac{3}{2},n}^{{HH}_1 \pm}=E_{-j_z\pm\frac{3}{2},n}^{{HH}_1 \pm}$ that allow for elastic scattering between these levels, thus precluding protection~\cite{spin-flip}. As shown in the SM (Sec.~IV), this picture still holds in the presence of spin-orbit and light--heavy-hole mixing effects; these lead to small ($\sim 1$ meV) energy level shifts comparable to the respective level broadenings, and hence can be neglected (this is also true for our Bi-based BHZ QDs, SM, Sec.~V).

\red{\paragraph*{Circulating current densities: $\boldsymbol{j}(\boldsymbol{r})$. ---}  We define $\boldsymbol{j}(\boldsymbol{r})=\frac{e\hbar}{m_{0}}Im\left\{\psi^{\dagger}(\boldsymbol{r})\nabla\psi(\boldsymbol{r})\right\}$,  where the total QD wave function $\psi(\boldsymbol{r})=\sum_{i} F_{i}(\boldsymbol{r})u_{i}(\boldsymbol{r})$ is  expressed as the sum of the product of the  periodic part of the Bloch function $u_{i}(\boldsymbol{r})$ of band $i$ at the $\Gamma$ point and its respective  envelope function $F_{i}(\boldsymbol{r})$.} \green{The average current over the unit cell is given by}~\cite{michael1,michael2} $\left\langle \boldsymbol{j}\right\rangle \left(\boldsymbol{r}\right)=\frac{e\hbar}{m_{0}}\thinspace Im\sum_{i,j}\left\{ F_{i}^{*}\left(\boldsymbol{r}\right)F_{j}\left(\boldsymbol{r}\right)\left\langle u_{i}\right|\boldsymbol{\nabla}\left|u_{j}\right\rangle +\delta_{ij}F_{i}^{*}\left(\boldsymbol{r}\right)\boldsymbol{\nabla}F_{j}\left(\boldsymbol{r}\right)\right\}$. Using \red{the wave function in Eq.~(\ref{eq6}) (see SM), we find}
\begin{eqnarray}
&\left\langle \boldsymbol{j}_{j_{z},n}^{\pm}\right\rangle  =\pm e\frac{N^{2}}{2\pi}\left\{ \frac{\sqrt{2}P}{\hbar}\thinspace\left|f_{1}^{\pm}\left(z\right)\right|\left|f_{3}^{\pm}\left(z\right)\right|I_{E_1,n}^{j_{z}\mp\frac{1}{2}}\left(r\right)I_{{HH}_1,n}^{j_{z}\mp\frac{3}{2}}\left(r\right) \nonumber \right. \\
 &\pm\frac{\hbar}{r\thinspace m_{0}}\left(j_{z}\mp\frac{1}{2}\right)\left[\left|f_{1}^{\pm}\left(z\right)\right|^{2}+\left|\thinspace f_{4}^{\pm}\left(z\right)\right|^{2}\right]\left|I_{E_{1},n}^{j_{z}\mp\frac{1}{2}}\left(r\right)\right|^{2} \nonumber \\
&\pm\frac{\hbar}{r\thinspace m_{0}}\left. \left(j_{z}\mp\frac{3}{2}\right)\left|f_{3}^{\pm}\left(z\right)\right|^{2}\left|I_{HH_1,n}^{j_{z}\mp\frac{3}{2}}\left(r\right)\right|^{2}\right\} \hat{\theta},
\label{currentfinal}
\end{eqnarray}
where the Kane parameter $P$ appears here due to coupling between conduction and valence bands. Here, the first term is the ``Bloch velocity" contribution to the average current as it stems from the periodic \red{part of the} Bloch function, while the second term is the contribution from the envelope function \cite{michael1,michael2}. Using $j_z\sim1$, $P=0.9055$~eV.nm and $r\sim R=40~$nm \red{we estimate the ratio of the Bloch to envelope contributions} $\left(\frac{\sqrt{2}P}{\hbar}\right)/\left(2\times\frac{\hbar}{Rm_0}\right)\sim 340$, \red{thus showing we can neglect the envelope velocity part in agreement with Ref.~\cite{michael2} (see SM, \green{Sec.~VIII} for a detailed comparison)}. \red{Since} $I_{E_1,n}^{j_{z}\mp\frac{1}{2}}=I_{E_1,n}^{-j_{z}\pm\frac{1}{2}}$ and $I_{{HH}_1,n}^{j_{z}\mp\frac{3}{2}}=I_{{HH}_1,n}^{-j_{z}\pm\frac{3}{2}}$, we find
\begin{equation}
\left\langle \boldsymbol{j}_{j_{z},n}^{\pm}\right\rangle \left(\boldsymbol{r}\right)=-\left\langle \boldsymbol{j}_{-j_{z},n}^{\mp}\right\rangle \left(\boldsymbol{r}\right),
\label{helical}
\end{equation}
\red{which shows the helical nature of the edge-like states within the gray region in Figs.~\ref{fig1}(c) and \ref{fig1}(e).}

To compare the topological QD edge states and the edge-like states in the trivial QD\red{, we plot Eq.~(\ref{currentfinal}) in Fig.~\ref{fig2}} for \red{the spin up QD} levels $j_z=\frac{3}{2}$ and $j_z=\frac{5}{2}$ \red{[see Figs.~\ref{fig1}(c) and \ref{fig1}(e), gray area] with $R=40$nm}.  \green{Interestingly, although the $j_z=3/2$ wave functions of both trivial and topological QDs  are extended, their circulating currents are localized near the QD edges. This arises from the product of the upper and lower wave function components in Eq.~(\ref{currentfinal}).} We find the highest current densities for the trivial edge-like states (due to the smaller $d$), Figs.~\ref{fig2}(a), \ref{fig2}(b). 
However, the integrated current density over half of the cross section of the QD $I_{j_z,n}^{\pm}=\int d\boldsymbol{S}\cdot \left\langle \boldsymbol{j}_{j_{z},n}^{\pm}\right\rangle=\int_{0}^{R}dr \int_{-\frac{d}{2}}^{\frac{d}{2}}dz \left|\left\langle \boldsymbol{j}_{j_{z},n}^{\pm}\right\rangle\right| \sim 0.17$ $\mu$A for both topological and trivial edge states to within 2$\%$, i.e., it shows no significant difference.

\paragraph*{Linear conductance. ---} To further compare \red{the} topological and trivial edge-like states, we calculate the two-terminal linear-response \red{QD} conductance ${\cal G}$ (at $T=0$K)~\green{\cite{meir-wingreen}} by coupling the \red{dots} to left (L) and right (R) leads, Fig.~\ref{fig1}(b).
Our Hamiltonian reads
\begin{equation}
\begin{aligned}{\cal H} & =\sum_{i}\varepsilon_{i}d_{i}^{\dagger}d_{i}+\sum_{k_{\alpha},\alpha,\sigma}\varepsilon_{k_{\alpha}\sigma}c_{k_{\alpha}\sigma}^{\dagger}c_{k_{\alpha}\sigma}+\sum_{i,k_{\alpha},\alpha,\sigma}V_{k_{\alpha}\sigma}^{i}d_{i}^{\dagger}c_{k_{\alpha}\sigma}\\
 & +\sum_{i\neq j}t_{ij}d_{j}^{\dagger}d_{i}+H.C., 
\end{aligned} \label{eq9}
\end{equation}
where $d_{i}^{\dagger}$ creates an electron in the QD state $\left|i\right\rangle$ [Eq.~(\ref{eq6})] \red{with energy $\varepsilon_i=\varepsilon_i(R,V_g)$ [obtained from Eq.~(\ref{eq5})], $i$ denotes the set of QD quantum numbers $j_z$, $\pm$ (or $\uparrow$,$\downarrow$~\cite{bhzspin}), and $n$ ($V_g$ is an additional gate controlling dot levels with respect to the Fermi energy of the leads), and $c_{k_{\alpha}\sigma}^{\dagger}$ creates an electron in the lead $\alpha=L,R$ with wave-vector $k_{\alpha}$, energy $\varepsilon_{k_{\alpha}\sigma}$ and  spin component $\sigma=~\uparrow,\downarrow$.}
\red{The spin-conserving matrix element $V_{k_{\alpha}\sigma}^{i}$ denotes the dot-lead coupling, while $t_{ij}$ couples the dot levels. }
\red{Next we focus on only four QDs states with well-defined $\sigma$, as illustrated in Fig.~\ref{fig3}(a)}. \red{This can be achieved by tuning the conduction window and the QD levels via external gates.}

\red{Figures~\ref{fig3}(b) and \ref{fig3}(c) show the QD conductance ${\cal G}={\cal G}_\uparrow + {\cal G}_\downarrow$ for the four  topological and trivial edge states with $j_{z}=\pm 9/2$ and $j_{z}=\pm 7/2$  [see \green{green triangles in} Figs.~\ref{fig1}(c) and \ref{fig1}(e)], as a function of the QD radius $R$ and the gate potential $V_g$.} The radius $R$ can be varied experimentally through an electrostatic confining potential~\cite{leon}. 
\red{The conductance for both the topological and trivial edge-like states show similar behaviors, i.e., double Lorentzian-like profiles centered at the QD levels $\varepsilon_i(R,V_g)$, separated by a dip, and peaked at $2e^2/h$}; this is clearly seen in the insets of Figs.~\ref{fig3}(b) and \ref{fig3}(c)] for two distinct $R$'s. \red{The dip follows from a destructive interference between the two same-spin edge states in the overlapping tails of the broadened QD density of states. See SM \green{(Sec.~IX)} where the conductance ${\cal G}$ is expressed as a sum of interfering amplitudes using Green functions~\cite{dip}.}


\red{Interestingly, bulk-like and edge-like valence edge states can coexist and even be degenerate in energy. In this case, our calculated conductances exhibit a crossover from a double-peak  resonance for $R<R_c$~nm and $V_g<V_{g,c}$ to a single-peak resonance at $R=R_c$~nm and $V_g=V_{g,c}$ and back to a double-peak resonance for $R>R_c$~nm and $V_g>V_{g,c}$. This is shown in Fig.~\ref{fig3}(e) (and its insets) for a trivial QD, but a similar plot also holds for a topological QD.  In the SM \green{(Sec. IX)} we show that when the bulk and edge-state Kramers pairs obey $\varepsilon_{3(4)}-\varepsilon_{1(2)}= t\thinspace\left(\frac{V^{3(4)}}{V^{1(2)}}-\frac{V^{1(2)}}{V^{3(4)}}\right)$, two of the transport channels are completely decoupled from the leads and hence a single resonance (peaked ${\cal G}=2 e^2/\hbar$) emerges. For the parameters in Fig.~\ref{fig3}(e) this decoupling occurs when the two Kramers pairs become degenerate, i.e., $\varepsilon_{1,2}(R_c,V_{g,c}) =\varepsilon_{3,4}(R_c,V_{g,c})$.}

\paragraph*{Concluding remarks.---} We have shown that Bi-based InAs QWs can become room-temperature TIs ($\sim 30$ meV) for well widths $d>6.9$ nm. Our realistic valence band anticrossing theory together with the $\boldsymbol{k.p}$ method allows us to calculate the  parameters of an effective BHZ model from which we can define cylindrical QDs via further confinement. By solving the BHZ QD eigenvalue equation analytically, we find protected helical edge states with equivalent circulating currents for both topological and non-topological regimes. Interestingly, we find that both topological and trivial QDs show similar transport properties, e.g., the two-terminal conductance ${\cal G}$ exhibits a two-peak resonance profile as a function of the QD radius and the gate $V_g$ controling its energy levels relative to the Fermi level of the leads. Hence from the point of view of two-terminal conductance probes, our proposed cylindrical QDs -- topological and non-topological -- are equivalent. We expect that our work stimulate experimental research on this topic.  

\begin{acknowledgments}
This work was supported by CNPq, CAPES, UFRN/MEC, FAPESP, PRP-USP/Q-NANO and the Center for Emergent Materials, a NSF MRSEC under Award No. DMR-1420451. We acknowledge the kind hospitality at the International Institute of Physics (IIP/UFRN), where part of this work was done and valuable discussions with Poliana H. Heiffig and Joost van Bree. DRC also acknowledges useful discussions with Edson Vernek and Lu\' \i s G. G. D. Silva.
\end{acknowledgments}


\begin{thebibliography}{30}
\bibitem{bhzspin} Due to the mixing between spin up $\uparrow$ and spin down $\downarrow$ components within the $\left|E_{1}\pm\right\rangle$ subbands, the spin index $\uparrow(\downarrow)$  is not a good quantum number within the BHZ subspace $+(-)$.  However, we calculate $\left\langle j_z\pm\right|s_{z}\left|j_z\pm\right\rangle \simeq\pm\frac{\hbar}{2}$; hence it is an excellent approximation to identify the $\pm$ subspace with $\uparrow,\downarrow$.
\end{thebibliography}
\end{document}